\documentclass[aps,prx,reprint,twocolumn, amsmath,amssymb,amsfonts,footinbib,
superscriptaddress,longbibliography,
notitlepage]{revtex4-2}

\AtBeginDocument{\usepackage{booktabs}}               
\makeatletter
\g@addto@macro\bfseries{\boldmath}
\makeatother

\usepackage[varg]{txfonts}
\usepackage[T1]{fontenc}
\usepackage[utf8]{inputenc}

\usepackage[varg]{txfonts}
\usepackage{hyperref}
\usepackage{amsmath}
\usepackage{color}
\usepackage{graphicx}
\usepackage[percent]{overpic}
\usepackage{mathrsfs}
\usepackage{bm}
\usepackage{braket}
\usepackage{mathtools}
\usepackage{bbold}
\usepackage{comment}
\usepackage[version=4]{mhchem}
\usepackage{xcolor}

\usepackage{soul,xcolor}
\setstcolor{red}
\definecolor{lightblue}{rgb}{.90,.95,1}
\definecolor{greeny}{rgb}{0.9,1,0}
\sethlcolor{greeny}
\definecolor{jaune}{RGB}{255,253,55}
\sethlcolor{jaune}

\definecolor{cred}{RGB}{228,26,28}
\definecolor{cblue}{RGB}{8,48,107}
\definecolor{cgreen}{RGB}{77,175,74}
\definecolor{cgray}{RGB}{150,150,150}
\definecolor{clgray}{RGB}{200,200,200}
\definecolor{cpurple}{RGB}{152,78,163}
\definecolor{corange}{RGB}{255,127,0}
\definecolor{cgold}{RGB}{230,171,2}

\newcommand{\avg}[1]{\langle #1 \rangle}

\newcommand{\meV}{{\ \rm meV}}
\newcommand{\hc}{{\rm h.c.}}
\newcommand{\subref}[2]{\ref{#1}\hyperref[#1]{#2}}

\renewcommand{\vec}[1]{\boldsymbol{#1}}

\newcommand{\vhat}[1]{\vec{\hat{#1}}}

\definecolor{cred}{RGB}{188,55,84}
\definecolor{cblue}{RGB}{55,126,184}
\hypersetup{colorlinks=true,linkcolor=cred,citecolor=cred,urlcolor=blue}

\graphicspath{{./figs/}}

\begin{document}
\title{Finite-Size Spectral Signatures of 
Order by Quantum Disorder: \\
A Perspective from Anderson's Tower of States}
\author{Subhankar Khatua}
\affiliation{Department of Physics and Astronomy, University of Waterloo, Waterloo, Ontario, N2L 3G1, Canada}
\affiliation{Department of Physics, University of Windsor, 401 Sunset Avenue, Windsor, Ontario, N9B 3P4, Canada}
\affiliation{
Institute for Theoretical Solid State Physics, IFW Dresden and W\"urzburg-Dresden Cluster of Excellence ctd.qmat, Helmholtzstrasse 20, 01069 Dresden, Germany}
\author{Griffin C. Howson}
\affiliation{Department of Physics and Astronomy, University of Waterloo, Waterloo, Ontario, N2L 3G1, Canada}
\affiliation{Department of Physics, University of Windsor, 401 Sunset Avenue, Windsor, Ontario, N9B 3P4, Canada}
\author{Michel J. P. Gingras}
\affiliation{Department of Physics and Astronomy, University of Waterloo, Waterloo, Ontario, N2L 3G1, Canada}
\affiliation{Waterloo Institute for Nanotechnology, University of Waterloo, Waterloo, Ontario, N2L 3G1, Canada}
\author{Jeffrey G. Rau}
\affiliation{Department of Physics, University of Windsor, 401 Sunset Avenue, Windsor, Ontario, N9B 3P4, Canada}

\newcommand{\g}{\textit{g}}
\newcommand{\uone}{U(1)}
\newcommand{\sutwo}{SU(2)}

\begin{abstract}
In frustrated magnetic systems with a subextensive number of classical ground states, quantum zero-point fluctuations can select a unique long-range ordered state, a celebrated phenomenon referred to as \emph{order by quantum disorder} (ObQD). 
For frustrated spin-$\frac{1}{2}$ models, unbiased numerical methods able to expose ObQD are necessary.
We show that ObQD can be identified from exact diagonalization (ED) calculations through an analysis akin to the Anderson tower of states associated with spontaneous symmetry breaking. 
By defining an effective quantum rotor model, we describe the competition between ObQD-induced localization of the rotor and its tunneling between symmetry-related ground states, identifying the crossover lengthscale from the finite-size regime where the rotor is delocalized, to the infinite system-size limit where it becomes localized. 
This rotor model relates the characteristic splittings in the ED energy spectrum to the ObQD selection energy scale, providing an estimate that can be compared to spin wave calculations. 
We demonstrate the general applicability of this approach in one-, two- and three-dimensional frustrated spin models that exhibit ObQD. 

\end{abstract}

\date{\today}

\maketitle
Many theoretical models of highly frustrated spin systems display an exponentially large manifold of accidentally degenerate classical ground states that is not a consequence of any global symmetry~\cite{Anderson1956,Villain1979,Chalker1992,Moessner1998b,Bramwell2001,Balents2010}.
This extensive degeneracy may give rise to a spin liquid phase in which the spins remain fluctuating down to zero temperature~\cite{Anderson1956,Moessner1998b,Bramwell2001,Balents2010,Gingras_McClarty,Savary_Balents}. 
However, real frustrated magnetic materials generally possess additional small interactions that can lift such accidental classical ground state degeneracy, yielding a long-range ordered state, albeit with residual quantum fluctuations, and thus forestalling the formation of a spin liquid state~\cite{Balents2010,Lacroix2011,Trebst2022}.

\begin{figure}[!h]
\centering
\includegraphics[width=0.9\columnwidth]{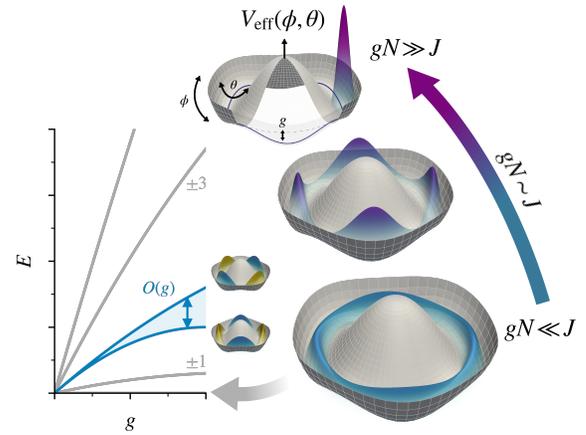}
\caption{Illustration of order by quantum disorder (ObQD) selection, with selection energy scale $\g$, showing the evolution from the finite-size limit $\g N \ll J$, to the thermodynamic limit $\g N\gg J$ (where $N$ is the number of spins). For small systems, tunneling delocalizes the order parameter across classical ground states, with ObQD acting as a small perturbation. (Inset) The scale of this perturbation can be read off the splittings of the excited states in the (approximate) Anderson tower of states. The order parameter localizes at the ObQD-selected state as $N\rightarrow \infty$.
}
\label{fig.cartoon}
\end{figure}
Interestingly, an intermediate case is possible where, despite perturbing interactions, some degree of frustration persists, with  the situation sitting between the two aforementioned extreme limits: instead of a unique ground state (up to global symmetries) or an exponentially large number of ground states, only a subextensive manifold of classical ground states remains. 
In such a scenario, quantum fluctuations may, perhaps counterintuitively, stabilize long-range magnetic order rather than prevent it---a phenomenon known as ``order by quantum disorder'' (ObQD)~\cite{Shender1982}.
ObQD has drawn significant attention over the years, with many spin models having been found to exhibit this phenomenon~\cite{Shender1982,Rastelli1987,Henley1989,tessman1954,Kubo1991,Chubukov1992,Henley1994,McClarty2014,Danu2016,Rau2018,Schick2020,Khatua2021,Noculak2023,Khatua2024,Hickey2024,Hickey2025} and some  potential experimental realizations of ObQD in materials~\cite{Brueckel1988,Kim1999,Champion2003,Zhitomirsky2012,Savary2012,Ross2014,Sarkis2020,Elliot2021}. 

Linear spin wave theory (LSWT)~\cite{Holstein1940,Auerbach1998} has been the prevailing method for studying ObQD in frustrated spin models~\cite{Shender1982,Rastelli1987,Henley1989,tessman1954,Kubo1991,Chubukov1992,Henley1994,McClarty2014,Danu2016,Rau2018,Schick2020,Khatua2021,Noculak2023,Khatua2024,Hickey2024,Brueckel1988,Kim1999,Champion2003,Zhitomirsky2012,Savary2012,Ross2014,Sarkis2020,Elliot2021,Hickey2025}.
This \emph{semiclassical} formalism treats quantum fluctuations as a perturbation about a classically ordered state stabilized for large spin length $S$.
The leading quantum correction to the energy is the zero-point energy (ZPE) of each classically degenerate ground state and the state with smallest ZPE is selected---order by quantum fluctuations (i.e.~disorder). 
It remains an open question to what extent this approach can be applied to systems with smaller, more realistic values of $S$. 
A quantitative understanding of ObQD in spin-$\frac{1}{2}$ systems, of foremost interest to experimentalists and theorists alike, requires a fully quantum approach, not rooted in the semiclassical limit. 
Since exact analytical treatments are scarce for frustrated $S=\frac{1}{2}$ systems, numerical techniques are essential. 
While there exist many such methods, exact diagonalization (ED) is often the tool of choice~\cite{Sandvik2010, Lauchli2011} as it is applicable to general spin models, provides access to the full spectrum, and can yield useful results in one, two, and even three dimensions.
Other numerical approaches~\cite{Sandvik1991,Syljuåsen2002, Kawashima2004,Assaad2022,White1992,SCHOLLWOCK201196,Verstraete2008,Rigol2006,Tang2013,Sandvik2010} fall short on one or more of these points, limiting their applicability in characterizing ObQD.

Unfortunately, being constrained to small system sizes, ED suffers from strong finite-size effects~\cite{Sandvik2010, Lauchli2011}. 
Notwithstanding, this limitation can reveal valuable insights since the finite-size low-energy spectrum contains structure that reflects the underlying ordering mechanism. 
This is exemplified by the ``Anderson tower of states'', a hallmark of spontaneous symmetry breaking (SSB) in finite size systems~\cite{Anderson1952,Anderson1997,Anderson2018,Lhuillier2005,Wietek2017}. 
The energy of the states in this tower scales inversely with system size (i.e.,~as $1/N$), collapsing into a degenerate set of states as $N \rightarrow \infty$, with the states that spontaneously break symmetry emerging as  superpositions within this degenerate set.
Despite its success in describing how SSB appears in finite-size spin-$\frac{1}{2}$ systems~\cite{Lhuillier2005,Wietek2017,Neuberger1989,Bernu1992,Hasenfratz1993,Azaria1993,Bernu1994,Fouet2001}, there is no similar understanding for ObQD~\cite{Lecheminant1995,Zhang2002,Lhuillier2005,Khatua2021}. This raises the fundamental and, to the best of our knowledge, unaddressed question---how does ObQD manifest in the finite-size energy spectrum of spin-$\frac{1}{2}$ systems and how is ObQD selection in the thermodynamic limit foreshadowed?

In this Letter, we identify signatures of ObQD in the low-energy spectrum of finite-size spin-$\frac{1}{2}$ systems and propose a prescriptive methodology for diagnosing ObQD in ED results. 
These signatures are encoded in a finite-size description of the dynamics of the ObQD order parameter, formulated in terms of an effective quantum rotor model that extends the conventional Anderson tower-of-states framework. 
This rotor model reproduces the spectroscopic signatures observed in ED data and captures the interplay between ObQD-induced localization of the rotor and quantum tunneling among symmetry-related ground states (see Fig.~\ref{fig.cartoon}).
In particular, this framework exposes the crossover from the finite-size regime, where the rotor is delocalized over the quasi-degenerate manifold, to the thermodynamic limit, where it localizes and long-range order occurs. 
By exploiting the characteristic splittings identified in the ED spectrum, we extract the ObQD energy scale and argue that these splittings constitute the finite-size manifestation of the energy gap of the ObQD-induced pseudo-Goldstone (PG) mode~\cite{Rau2018}. 
We compare estimates of the ObQD energy scale from these $S=\frac{1}{2}$ calculations with semiclassical results for several paradigmatic ObQD models and find reasonable agreement.  
Finally, we discuss implications for the pyrochlore magnet \ce{Er2Ti2O7}, a leading candidate material for ObQD~\cite{Savary2012,Zhitomirsky2012,Ross2014}.
 
\begin{figure}
\includegraphics[width=\columnwidth]{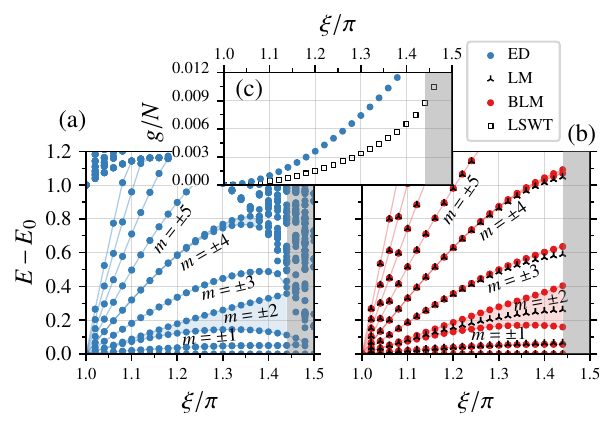}
\caption{Ferromagnetic Heisenberg-compass model on the square lattice of $N = 16$ sites with periodic boundary conditions. 
(a) Low-energy ED spectrum. The $m$ labels correspond to the unsplit rotor energy levels, with the splitting of the $m = \pm 2$ and $m = \pm 4$ levels emphasized by light blue shaded wedges. The lines are guides to the eye. 
(b) Spectrum of the LM and BLM models, where the $\g$ value for the BLM description is extracted from the splitting in the ED spectrum.
(c) Comparison of the ObQD energy scale, $\g$, obtained from finite-size LSWT for $N=16$ and ED data. 
Gray areas in all panels mark deviations from the BLM picture due to a new phase near $\xi =3\pi/2$~\cite{Khatua2024}.} 
\label{fig.HC-FM-spec}
\end{figure}

{\it{Ferromagnetic Heisenberg-compass model.}}--- 
We first consider the ferromagnetic Heisenberg-compass model on the square lattice~\cite{Khatua2024,Khatua2023}, relevant to spin-orbit Mott insulators~\cite{Nussinov-compass-rev}
\begin{equation}
  {\cal H} = \sum_{\vec{r}}\biggl[J\!\!\sum_{\vec{\delta} = \vec{x},\vec{y}} 
    \vec{S}^{\phantom{x}}_{\vec{r}} \!\cdot \vec{S}^{\phantom{x}}_{\vec{r}+\vec{\delta}\phantom{+\vec{}}}
    \!\!+K \left(S^x_{\vec{r}\phantom{+\vec{}}}\!\! S^x_{\vec{r}+\vec{x}} + S^y_{\vec{r}\phantom{+\vec{}}} \!\!S^y_{\vec{r}+\vec{y}}
  \right)\biggr],
  \label{eq.Ham-Hc}
\end{equation}
where $ \vec{S}_{\vec{r}}\equiv \left(S_{\vec{r}}^x, S_{\vec{r}}^y, S_{\vec{r}}^z\right)$ is a spin-$\frac{1}{2}$ operator at lattice site $\vec{r}$, and $\cramped{\vec{\delta} = \vec{x}, \vec{y}}$ are the nearest-neighbor bonds. 
The first and second terms in Eq.~\eqref{eq.Ham-Hc} are the nearest-neighbor Heisenberg and bond-dependent compass exchanges, respectively.
We parametrize $\cramped{J \equiv \cos\xi}$ and $\cramped{K \equiv \sin\xi}$ with $\cramped{\pi<\xi<3\pi/2}$ where both couplings are ferromagnetic (i.e., negative). 
The compass term breaks the \sutwo{} spin-rotation symmetry of the Heisenberg term, reducing the symmetry to discrete $C_{4z}$ and $C_{2x}$, $C_{2y}$ operations.

ObQD in this model has been previously studied using LSWT~\cite{Rau2018,Khatua2023,Khatua2024}.
In the classical limit $\cramped{(S\rightarrow\infty)}$, the ground states are ferromagnetic configurations aligned along \emph{any} direction in the $\cramped{\vhat{x}-\vhat{y}}$ plane, forming an accidentally degenerate continuous O(2) manifold.
Spin-wave zero-point fluctuations lift this degeneracy, yielding ObQD for states aligned along $\cramped{\pm\vhat{x}, \pm\vhat{y}}$ directions~\cite{Khatua2024,Rau2018}. 
To assess the extent to which the semiclassical understanding of ObQD carries over to the $S=\frac{1}{2}$ limit, we perform ED, exploiting the discrete translational symmetry of the model to obtain the low-energy spectrum shown in Fig.~\subref{fig.HC-FM-spec}{(a)}.
Since for small system sizes, the ED ground state shows only a very weak preference for ObQD-selected orderings~\cite{Khatua2024}, we look beyond the ground state to the excited states.

Do the excited states reveal signatures of ObQD? 
Specifically, is there any underlying structure in the spectrum connected to the ObQD observed in the thermodynamic limit? 
This question is reminiscent of the related challenge in identifying SSB in finite-size systems~\cite{Anderson1952,Lhuillier2005,Wietek2017}. 
To proceed, we formulate an effective description of ObQD based on the Lieb-Mattis (LM) framework~\cite{Lieb-Mattis}, previously used to understand finite-size signatures of SSB---namely, the Anderson tower of states~\cite{Anderson1952,Anderson1997,Anderson2018,Lhuillier2005,Wietek2017}. 
The LM Hamiltonian can be derived by projecting the full Hamiltonian [Eq.~\eqref{eq.Ham-Hc}] into its lowest-energy subspace~\cite{Zhang2002,Roscilde2023}: the ferromagnetic sector with total spin $S_{\rm tot} = \frac{N}{2}$, where $N$ is the total number of spins.
This yields an effective Hamiltonian in terms of a single large spin of length $S_{\rm tot}$: ${\cal H}_{\rm LM}=-K(S_{\rm tot}^z)^2/(N-1)$~\footnote{The $(N-1)$ denominator comes from the total number of interaction pairs per spin in this fully symmetric (maximum-spin) subspace.} where $S_{\rm tot}^z$ has eigenvalues $m = 0,\pm 1,\cdots, \pm S_{\rm tot}$~\cite{SM}.
This is a planar quantum rotor with moment of inertia $I = {(N-1)}/(2|K|)$ that increases with system size.  
Since $K<0$, the ground state has $m=0$, while the excited states are each doubly degenerate with $|m|=1,2,\dots$, $S_{\rm tot}$ [see Fig.~\subref{fig.HC-FM-spec}{(b)}]. 

\begin{figure}
\includegraphics[width=0.9\columnwidth]{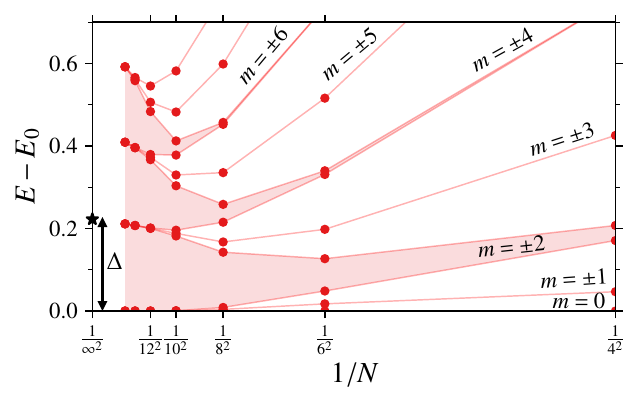}
\caption{Finite size scaling of the BLM spectrum of the Heisenberg-compass model with $\xi = 5\pi/4$. The $\g$ value used for the BLM model is determined from LSWT on an $N$-site lattice. Lines are guides to the eye, tracking the scaling of the energy eigenvalues. Shaded areas show ObQD-induced splitting. Black star marker is the system size extrapolated value of the gap $\Delta$ from the ground state to the first excited state reproducing the pseudo-Goldstone gap~\cite{Rau2018}.}
\label{fig.HC-FM-scaling}
\end{figure}
Comparing this with the ED spectrum in Fig.~\subref{fig.HC-FM-spec}{(a)}, we find a good qualitative agreement, \emph{except} for splittings visible in several excited doublets as $\xi$ (i.e., $|K|$) increases. 
While the LM rotor predicts doublets for the $m = \pm 2$ and $m = \pm 4$ energy levels, the ED data in Fig.~\subref{fig.HC-FM-spec}{(a)} shows this degeneracy is lifted. 
This discrepancy arises because ${\cal H}_{\rm LM}$ possesses a \uone{} symmetry about the $\vhat{z}$ axis while the microscopic spin Hamiltonian [Eq.~\eqref{eq.Ham-Hc}] does not. 
This symmetry of ${\cal H}_{\rm LM}$ reflects the accidental degeneracy of the classical ground states, and despite forming the foundation of the effective theory of SSB, ${\cal H}_{\rm LM}$ is unable to capture ObQD.

{\it{Beyond the Lieb-Mattis model.}}---  
To account for ObQD, it is thus necessary to go beyond the LM projection by breaking the emergent \uone{} symmetry and restoring the discrete symmetries of the full Hamiltonian~\cite{Zhang2002}. 
Such corrections can in principle be derived via second-order perturbation theory starting from the isotropic LM limit~\cite{Zhang2002}, 
however such calculations are challenging for Hamiltonians without continuous spin-rotation symmetry~\cite{Liviotti2002}. 

We therefore take a simpler alternative route: we introduce the \uone{} breaking term phenomenologically, inspired by spin-wave theory. 
In LSWT, the term that restores the discrete symmetries is the spin-wave ZPE which, at the leading order, is $-\g\cos(4\phi)$~\cite{Khatua2024}, where $\phi$ is the moment orientation of the ferromagnetic ground states relative to $\vhat{x}$ and the parameter $\g$ is the ObQD energy scale.  
We thus promote this ZPE to an operator~\cite{SM} 
\begin{equation}\label{eq.HC-FM-Hpert}
    {\cal H}_{\rm pert} = -\frac{\g}{2}\frac{\left[\big({S}_{\rm tot}^+\big)^4 + \big({S}_{\rm tot}^-\big)^4\right]}{\left[N/2(N/2 + 1)\right]^2},
\end{equation}
where $S_{\rm tot}^{\pm}$ are the raising and lowering operators acting within the fixed $\cramped{S_{\rm tot} = \frac{N}{2}}$ multiplet, increasing or decreasing the $S_{\rm tot}^z$ eigenvalue by one, and the denominator accounts for the length of $\vec{S}_{\rm tot}$, i.e., $\sqrt{N/2(N/2+1)}$~\footnote{The factor $[N/2(N/2+1)]^2$ is a phenomenological normalization constant. The true $N$ dependence could in principle be found from high-order many-body perturbation theory starting from the Heisenberg limit. This is computationally challenging for anisotropic Hamiltonians~\cite{Liviotti2002} and beyond the scope of this work.}. 
This minimal term embeds ObQD in the effective description, with $\g$ a phenomenological parameter setting the energy scale of the ground state selection~\footnote{For the same reason that this accidental degeneracy cannot be lifted classically by two-spin interaction terms, the correction to $\mathcal{H}_{\rm LM}$ needed to explain the ObQD splittings must be a higher-spin operator}.

The resulting ``beyond Lieb-Mattis'' (BLM) Hamiltonian, ${\cal H}_{\rm BLM} \equiv {\cal H}_{\rm LM} + {\cal H}_{\rm pert}$, describes the motion of the rotor in an ObQD-induced ``potential'', ${\cal H}_{\rm pert}$, with minima along $\pm\vhat{x}, \pm\vhat{y}$ directions.  
The effect of this potential on the spectrum of ${\cal H}_{\rm BLM}$ can be understood by treating ${\cal H}_{\rm pert}$ as a perturbation to ${\cal H}_{\rm LM}$, since $\g$ is taken to be an extensive quantity (i.e., $\propto N$) and therefore weaker for small system sizes. 
Due to the quartic dependence of ${\cal H}_{\rm pert}$ on ${S}_{\rm tot}^{\pm}$, degenerate states with $\Delta m = \pm 4$ split at first order in $\g$, 
while those with $\Delta m = \pm 8$ split at second order.
The $m = \pm 1, \pm 3, \dots$ states thus remain degenerate, while the $m = \pm 2$ and $m = \pm 4$ states exhibit splittings of $O(\g)$ and $O(\g^2)$, respectively---the latter being much smaller. 

These splittings relate the BLM and ED spectra. Focusing on the $m = \pm 2$ states, the BLM splitting, $\Delta$, is given perturbatively by
$\Delta = \g\frac{(N/2-1)(N/2+2)}{N/2(N/2+1)}$~\cite{SM}.
We can then use the same splitting from the ED spectrum, $\Delta_{\rm ED}$, and equate $\Delta = \Delta_{\rm ED}$ to produce an estimate for the value of $\g$ from the microscopic model.
A comparison of the ED spectrum [Fig.~\subref{fig.HC-FM-spec}{(a)}] and BLM spectrum using the extracted values of $\g$ [Fig.~\subref{fig.HC-FM-spec}{(b)}] reveals compelling qualitative agreement~\footnote{The disagreement near $\xi \sim 1.5\pi$ is due to a nearby competing phase~\cite{Khatua2024} rendering the effective BLM description inapplicable.}. 
This BLM description exposes that the pattern of splittings observed in the excited energy levels in the ED spectrum \emph{originates} from the physics of ObQD, further providing an estimate of the ObQD energy scale---this is the key result of this paper. 
We note that the ED spectrum deviates qualitatively from the BLM model predictions near $\xi \sim 3\pi/2$, marked by gray areas in all panels of Fig.~\ref{fig.HC-FM-spec}, where a large number of states appear at low energies. This disagreement arises because the system enters a new phase around $\xi\sim 3\pi/2 $, a nematic ordering along $\vhat{x}$ or $\vhat{y}$~\cite{Khatua2024}.

As the energy scale $\g$ was motivated by the form of the ObQD potential that appears in LSWT, it can be directly compared to the semiclassical value computed as half of the ZPE difference between the $\phi = \frac{\pi}{4}$ and $\phi =0$ states (see the Supplemental Material~\cite{SM}). As shown in Fig.~\subref{fig.HC-FM-spec}{(c)}, there is qualitative agreement---about a factor of two---between the value of $\g$ extracted from ED, $\g_{\rm ED}$, and the value of $\g$ directly computed from LSWT, $\g_{\rm LSWT}$. 
Additional ED results for larger system sizes, presented in the Supplemental Material~\cite{SM}, demonstrate that $\g$ extracted from this finite-size spin-$1/2$ spectroscopic analysis consistently approaches its semiclassical thermodynamic-limit value as the system size increases.

\begin{figure}
\includegraphics[width=\columnwidth]{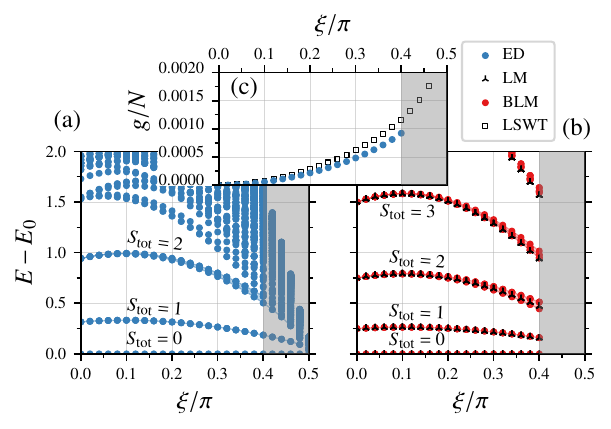}
\caption{Antiferromagnetic Heisenberg-Kitaev model on the honeycomb lattice of $N = 24$ sites with periodic boundary conditions. a) Low-energy ED spectrum. (b) Spectra of the LM and BLM models, where the $\g$ value for the BLM description is extracted from the splitting in ED spectrum. (c) $\g$ obtained from LSWT and the ED spectrum. Gray areas in all panels mark the regime where the low-energy physics begins to deviate from the BLM description as it approaches the Kitaev quantum spin liquid phase near $\xi=\pi/2$
~\cite{Chaloupka2013}.
}
\label{fig.HK-AFM-spec}
\end{figure}

The BLM Hamiltonian has two important limits: for smaller systems, the kinetic rotor term (${\cal H}_{\rm LM}$) dominates over the ObQD-induced potential as its moment of inertia scales $\propto (N-1)$; the rotor remains delocalized in the $\vhat{x}-\vhat{y}$ plane, hybridizing among the accidentally degenerate ground states. 
In the opposite limit of larger systems, the potential dominates, localizing the rotor near the potential minima and suppressing the tunneling through the potential barriers (see Fig.~\ref{fig.cartoon}). 
The crossover between these two limits can be seen in the finite-size scaling of the BLM spectrum at a fixed $\xi$, as shown in Fig.~\ref{fig.HC-FM-scaling} for a BLM model where, here, we have set $\g = \g_{\rm LSWT}$.
Groups of four states converge as $N$ increases, becoming quasi-degenerate for sufficiently large systems, consistent with the semiclassical picture of ObQD selecting four states in the thermodynamic limit. 
The near-degeneracy of the lowest four states at $N^* \approx 64$ marks the crossover from rotor-dominated to ObQD-dominated physics [see Fig.~\ref{fig.cartoon}].  
In the large-$N$ limit, the energy gap $\Delta$ to the first excited state shown in Fig.~\ref{fig.HC-FM-scaling} approaches the ObQD-induced PG gap determined from LSWT~\cite{Rau2018}.
The precise agreement of the PG gap from the two approaches in the large-$N$ limit is by construction, due to $\g$ being fixed using LSWT. 
While, to the best of our knowledge, no unbiased large-system-size numerical studies, e.g., quantum Monte Carlo (QMC) or density matrix renormalization group (DMRG), are available to directly compare with the gap calculated here, previous QMC studies in the compass limit $(\xi = 3\pi/2)$~\cite{Wenzel2008, Wenzel2010} suggests that such simulations may be feasible throughout the range $\pi<\xi<3\pi/2$, allowing the gap $\Delta$ to be computed. 
Thus, the BLM description offers a clear picture of the mechanism by which ObQD operates in spin-$\frac{1}{2}$ systems, from finite-size clusters to the thermodynamic limit. 

Interestingly, this Heisenberg-compass model also exhibits ObQD in the antiferromagnetic regime (i.e., $J>0$, $K>0$), but now into N\'eel order along the $\pm\vhat{x},\pm\vhat{y}$ directions~\cite{Khatua2024}. 
A direct application of the BLM framework yields a similar effective rotor physics in the finite-size spin-$\frac{1}{2}$ limit (see the Supplemental Material~\cite{SM}).

{\it{Antiferromagnetic Heisenberg-Kitaev model}.}--- 
To demonstrate the broad applicability of this framework, we consider another model, the Heisenberg-Kitaev model on the honeycomb lattice~\cite{Chaloupka2010,Chaloupka2013,Rau2016} relevant to candidate Kitaev materials~\cite{Chaloupka2010,Chaloupka2013,Trebst2022}, that yields qualitatively different rotor physics.  
The Hamiltonian is
$\cramped{{\cal H} ~=~ \sum_{\langle\vec{r}\vec{r}'\rangle\in \gamma}\left[J \vec{S}_{\vec{r}}\cdot\vec{S}_{\vec{r}'}+K S^\gamma_{\vec{r}} S^\gamma_{\vec{r}'}\right]},$
where $J, K$ are the Heisenberg and Kitaev couplings, respectively, and $\gamma = x, y, z$ labels the three types of nearest-neighbor bonds. 
The bond-dependent Kitaev interaction breaks \sutwo{} spin-rotation symmetry of the Heisenberg term down to discrete symmetries~\cite{SM}.

For $J>0$ and $K>0$, classical ground states are accidentally degenerate N\'eel states with arbitrary orientation, and ObQD selects those along $\pm \vhat{x}$, $\pm\vhat{y}$, $\pm\vhat{z}$ directions~\cite{Chaloupka2013,Chaloupka2010,Rau2018}. 
The low-energy $\cramped{S=\frac{1}{2}}$ ED spectrum is shown in Fig.~\subref{fig.HK-AFM-spec}{(a)} with $J \equiv \cos\xi$ and $K \equiv \sin\xi$ for $0<\xi<\pi/2$. 
To identify signatures of ObQD in this spectrum, we follow the same procedure as laid out above for the ferromagnetic Heisenberg-compass model.

Unlike an effective spin of length $N/2$ describing the ferromagnetic subspace, the relevant low-energy subspace now consists of two large spins of length $N/4$, corresponding to the two sublattices $A$ and $B$ of the N\'eel ordering, each containing $N/2$ spins. 
The resulting LM Hamiltonian is ${\cal H}_{\rm LM} = \frac{3J+K}{N} \big(\vec{S}_{\rm tot}^2 - \vec{S}^2_A - \vec{S}^2_B\big)$~\cite{SM}, where $\vec{S}_{\rm tot} = \vec{S}_{\rm A} + \vec{S}_{\rm B}$. 
Since ${\cal H}_{\rm LM}$ has \sutwo{} symmetry, $S_{\rm tot}$ and $S_{\rm tot}^z$ are conserved.
The ground state has $S_{\rm tot} = 0$, while excited states are labeled by $S_{\rm tot} = 1, 2, \dots$, each with degeneracy $(2S_{\rm tot}+1)$ [see Fig.~\subref{fig.HK-AFM-spec}{(b)}]. 
This qualitatively agrees with the ED spectrum [Fig.~\subref{fig.HK-AFM-spec}{(a)}], \emph{except} for the splitting of the $S_{\rm tot} = 2$ level and above observed in the ED spectrum. 
 
To resolve this discrepancy, we consider a BLM description, adding the operator equivalent of the spin-wave ZPE, as done previously.
Here, the ZPE manifests as a cubic anisotropy, with minima along the $\pm \vhat{x}$, $\pm\vhat{y}$, $\pm\vhat{z}$ directions. 
The corresponding operator is~\cite{SM}
\begin{equation}\label{eq.HK-honeycomb-pert}
{\cal H}_{\rm pert} = -\g \frac{(S_A^x - S_B^x)^4 + (S_A^y - S_B^y)^4 + (S_A^z - S_B^z)^4}{\left[N/4 (N/4 + 1)\right]^2},
\end{equation} 
where the denominator arises from the length of each effective large spin.  
The perturbative action of ${\cal H}_{\rm pert}$ on ${\cal H}_{\rm LM}$ mirrors the problem of a magnetic ion placed in a cubic crystal field environment~\cite{Fazekas1999}. 
It does not lift the degeneracy of the $S_{\rm tot} = 1$ level but splits the $S_{\rm tot} = 2$ level into the analogue of the $e_g$ doublet and $t_{2g}$ triplet~\cite{Fazekas1999}, 
as clearly observed in both the ED [Fig.~\subref{fig.HK-AFM-spec}{(a)}] and BLM spectra [Fig.~\subref{fig.HK-AFM-spec}{(b)}]. 
Moreover, $\g$ from LSWT~\cite{SM} agrees well with that extracted from the $S_{\rm tot} = 2$ splitting in the ED spectrum [Fig.~\subref{fig.HK-AFM-spec}{(c)}].
The disagreement between the BLM and ED spectra near $\xi = \pi/2$, marked by a gray region in all panels of Fig.~\ref{fig.HK-AFM-spec}, arises due to the presence of the Kitaev spin liquid phase around this point~\cite{Chaloupka2013}. 
The ferromagnetic counterpart of this model (i.e., $J<0$, $K<0$) also exhibits ObQD~\cite{Chaloupka2010,Chaloupka2013}, with the BLM framework again yielding an analogous rotor description in the finite-size spin-$\frac{1}{2}$ limit (details in the Supplemental Material~\cite{SM}). 
A one-dimensional version of the ferromagnetic Heisenberg-Kitaev model also exhibits ObQD~\cite{Yang2024}, and its effective rotor physics can be obtained by a straightforward application of the BLM framework, as discussed in the Supplemental Material~\cite{SM}.

{\it{Conclusion}.}---
To date, an understanding of the mechanism for order by quantum disorder (ObQD) has been lacking in finite-size $S=\frac{1}{2}$ systems.
We have considered two textbook models harboring ObQD and presented an effective beyond Lieb-Mattis (BLM) description of how ObQD operates in the spin-$\frac{1}{2}$ limit, from finite-size systems to the thermodynamic limit.  
We identified a pattern of splittings in the \emph{excited states} in the low-energy exact diagonalization (ED) spectrum, serving as a precursor to ObQD in the thermodynamic limit. 
Just as the Anderson tower of states serves as a diagnostic of spontaneous symmetry breaking in finite-size spin-$\frac{1}{2}$ systems, our analysis plays a similar role for ObQD.

The effective BLM description captures the crossover from a delocalized rotor in small systems to ObQD-dominated localized rotor in larger systems. 
This crossover can occur at system sizes significantly larger than those typically accessible to most unbiased numerical techniques. 
For example, in the ferromagnetic Heisenberg-compass model on the square lattice, it appears around $N^* \approx 64$. 
The value of $N^*$ depends on the model and may be much larger in systems with weaker ObQD-selection effects. 
Only if ED calculations were feasible for such sizes would one observe a clear signature of ObQD-induced ordering in the ground state. 

Our effective framework is general, applicable to any model exhibiting ObQD in one, two, or three dimensions, with no restrictions on symmetry or lattice type. 
We have demonstrated this generality via canonical two-dimensional examples, as well as a one-dimensional example~\cite{SM}.
We have also applied our framework to a three-dimensional case, $\ce{Er2Ti2O7}$, a preeminent candidate material for ObQD (see End Matter~\ref{app.ETO}). 
This pyrochlore magnet of effective spins $(S=\frac{1}{2})$ orders into a non-collinear ``$\psi_2$'' state~\cite{Savary2012,Ross2014}. 
We find signatures of ObQD in its ED spectrum, with qualitative agreement between the ObQD energy scale $\g$ extracted from the ED splittings and LSWT. 

While the LM and BLM descriptions are demonstrated in this Letter for spin-$\frac{1}{2}$ systems, which was the key motivation stated at the outset, they can also be applied to higher-spin systems. However, the formulation requires slight modifications: in deriving the LM part of the effective Hamiltonian, we have explicitly used identities specific to spin-$\frac{1}{2}$ Pauli matrices~\cite{SM}, which do not hold for higher spins and would lead to additional terms in the effective Hamiltonian. We expect these changes to modify the Hamiltonian quantitatively, while leaving the qualitative rotor interpretation unchanged.
 
The finite-size spectroscopic analysis of ObQD presented in this work reveals not only the selected ordering, but also the strength of the selection, which is roughly within a factor of two of its value estimated semi-classically in the thermodynamic limit for all the models we examined.   
Similar finite-size spectroscopic approaches have also proven successful in exposing the physics arising in the thermodynamic limit in other contexts including valence bond solid (VBS)~\cite{Lhuillier2005}, quantum criticality~\cite{Schuler2016,Whitsitt2016,Whitsitt2017,Schuler2021,Wietek2024} and quantum spin liquids (QSL)~\cite{Schuler2016,Whitsitt2016,Wietek2024}.  
Note that the BLM framework is applicable only to systems exhibiting magnetic long-range order, and is therefore not suitable for phases such as VBS or QSL, which preserve full spin-rotation symmetry. 
We believe our work significantly deepens the understanding of order by disorder in finite-size systems and in the extreme $S=\frac{1}{2}$ quantum limit. 
We hope it encourages further exploration of finite-size quantum spectra as a diagnostic tool for uncovering emergent order, and inspires new directions in the study of fluctuation-driven phenomena in frustrated quantum magnets.

\emph{Acknowledgments}---We thank F. F. Assaad, A. Hickey, A. L\"auchli, N. Perkins, I. Rousochatzakis, and J. van den Brink for discussions. 
The work at the University of Waterloo and at the University of Windsor was funded by the NSERC of Canada~(M.J.P.G, J.G.R) and the Canada Research Chair Program~(M.J.P.G, Tier I). 
S. K. acknowledges financial support from the Deutsche Forschungsgemeinschaft (DFG, German Research Foundation) under Germany’s Excellence Strategy through the Hallwachs-R\"ontgen Postdoc Program of the W\"urzburg-Dresden Cluster of Excellence ctd.qmat---Complexity, Topology and Dynamics in Quantum Matter (EXC 2147, Project-ID 390858490) and thanks Ulrike Nitzsche for technical assistance.
G.C.H. acknowledges funding from the NSERC of Canada through the USRA and CGS-M programs.
This research was supported in part by grant NSF PHY-1748958 to the Kavli Institute for Theoretical Physics (KITP).

\bibliography{main}

\onecolumngrid
\clearpage
\section*{End Matter}
\twocolumngrid

\renewcommand{\theequation}{E\arabic{equation}}
\renewcommand{\thefigure}{E\arabic{figure}}
\renewcommand{\thetable}{E\arabic{table}}
\setcounter{equation}{0}
\setcounter{figure}{0}
\setcounter{table}{0}

\appendix
\section{Application to \ce{Er2Ti2O7}}
\label{app.ETO}

As another non-trivial application of the above framework, we consider one of the best material examples of ObQD, \ce{Er2Ti2O7}~\cite{Savary2012,Zhitomirsky2012}.
The magnetic physics of this compound is described by an effective spin-$\frac{1}{2}$ model for the lowest-lying crystal field doublet of the \ce{Er^{3+}} ion~\cite{Rau2019}. At nearest neighbor level symmetry constrains this to four independent exchanges~\cite{Rau2019}
\begin{align}
\label{eq:model}
{\cal H} &\equiv \sum_{\avg{ij}} \left[
J_{zz} S^z_i S^z_j - J_{\pm}\left(S^+_i S^-_j+S^-_i S^+_j\right)+\right. \nonumber \\
& J_{\pm\pm} \left(\gamma_{ij} S^+_i S^+_j+\hc \right)+ 
 \left. J_{z\pm}  \left(
\zeta_{ij} \left[ S^z_i S^+_j+ S^+_i S^z_j \right]+ \hc \right)\right].
\end{align}
Fitting to inelastic neutron scattering experiments yields the parameters~\cite{Savary2012}
\begin{align*}
J_{zz} &= -0.025 \meV, &	J_{\pm} &= +0.065 \meV ,\\	
J_{\pm\pm} &= +0.042 \meV, & 	J_{z\pm}  &= -0.0088 \meV.
\end{align*}
Classically, a ferromagnetic $XY$ phase is stabilized, $\vec{S}_i \equiv S\left(\vhat{x} \cos{\phi} +  \vhat{y} \sin{\phi}\right)$, with an accidental \uone{} degeneracy in the angle $\phi$. 
Quantum fluctuations lift this degeneracy and select the $\psi_2$ state, corresponding in this language to $\phi = 2\pi n/6$ with $n=0,1,\dots,5$~\cite{Savary2012,Zhitomirsky2012}. The ObQD potential can be modeled as $V(\phi) = -\g\cos{(6\phi)}$ due to the three-fold symmetry of the lattice~\cite{Savary2012}.

\begin{figure}[t]
    \centering
    \includegraphics[width=\linewidth]{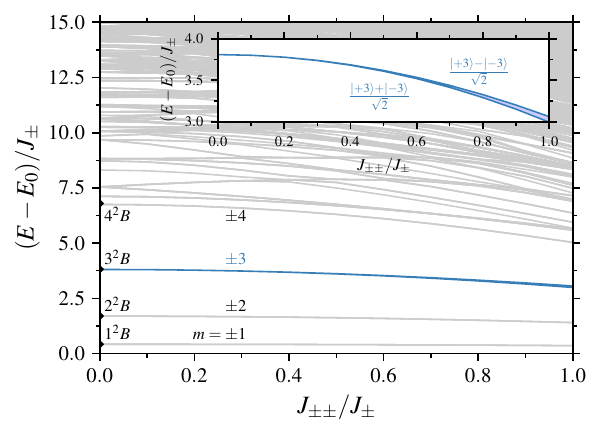}
    \caption{Spectrum of a single cubic unit cell ($N=16$) of the best fit model Eq.~(\ref{eq:model}) of \ce{Er2Ti2O7} tuning away from the $XY$ limit.
    Low-lying states form an approximate tower corresponding to a \uone{} rotor with $(E-E_0)/J_{\pm} = m^2 B$ where $m=0,\pm1,\pm2,\dots$ and $B\approx 0.4248$. (Inset) Splitting of the $m=\pm 3$  levels into even and odd combinations $(\ket{+3}\pm \ket{-3})/\sqrt{2}$ is a direct indicator of the ObQD strength.
    }
    \label{fig:eto-spectrum}
\end{figure}

To simplify our analysis we will consider a minimal version of this  model with $J_{zz} = J_{z\pm}=0$ and vary the ratio $J_{\pm\pm}/J_{\pm}$~\cite{Zhitomirsky2012}. We have performed full ED on a single cubic unit cell of the pyrochlore lattice of $N=16$ spins, exploiting translation invariance to reduce the block size. The spectrum as a function of $J_{\pm\pm}/J_{\pm}$ is shown in Fig.~\ref{fig:eto-spectrum}. The low lying states form an Anderson tower corresponding to a \uone{} quantum rotor, with energy levels approximately given by $m^2 B J_{\pm}$ where $m$ is an integer and $B\approx 0.4248$. 

\begin{figure}[!b]
    \centering
    \includegraphics[width=\linewidth]{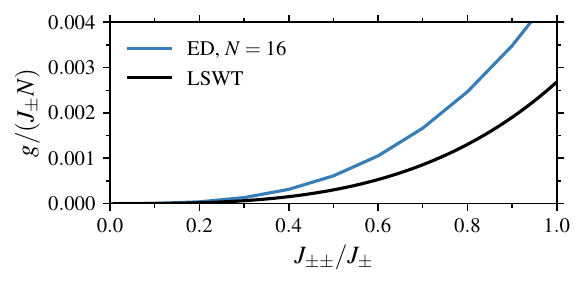}
    \caption{Splitting of the $m = \pm 3$ level in ED (see Fig.~\ref{fig:eto-spectrum}) yields the coefficient, $\g$, in the ObQD selection energy $-\g\cos(6\phi)$; corresponding semiclassical estimate from LSWT is shown.
    }
    \label{fig:eto-splitting}
\end{figure}

We can see that there is a very weak splitting in the third excited doublet, corresponding to mixing of the $m=\pm 3$ states. 
Explicitly, following the methodology presented in the main text, we construct a BLM model of the form
$$
\mathcal{H}_{\rm BLM} = B (S^z_{\rm tot})^2
-\frac{\g}{2} \left\{\frac{(S^+_{\rm tot})^6+(S^-_{\rm tot})^6}{[N/2(N/2+1)]^3}\right\},
$$
where $S_{\rm tot}^z = \sum_i S^z_i$ and $N=16$. Following the same line of argument as the main text, we can extract the coefficient $\g$ directly from the splitting,
with the splitting given as
$$
\Delta_{\rm BLM} = \g\left\{\frac{(N/2+3)(N/2+2)(N/2-1)(N/2-2)}{[N/2(N/2+1)]^2}\right\}.
$$
For a sixteen site cluster this implies that $\g \approx 1.122 \Delta_{\rm BLM}$.
The value of $\g$ obtained using the splitting from ED spectrum as a function of $J_{\pm\pm}/J_{\pm}$ is shown in Fig.~\ref{fig:eto-splitting}, alongside the value for $\g$ inferred from LSWT by comparison of the ZPEs of the $\psi_2$ and $\psi_3$ states~\cite{Savary2012}.

Qualitative agreement is seen even for this small size, with the ED results larger than the LSWT by a factor of approximately two. Neglecting $J_{zz}$ and $J_{z\pm}$ the best fit exchanges for \ce{Er2Ti2O7} would fall at $J_{\pm\pm}/J_{\pm} \approx 0.646$ with $\g_{\rm ED} \approx 0.0875\ \mu{\rm eV}$ and 
$\g_{\rm LSWT} \approx 0.0436\ \mu{\rm eV}$. These are comparable to the values extracted for the full set of exchanges, including $J_{zz}$ and $J_{z\pm}$, which yields
$\g_{\rm ED} \approx 0.0797\ \mu{\rm eV}$ and 
$\g_{\rm LSWT} \approx 0.0423\ \mu{\rm eV}$.

We note that calculations based on a real-space perturbation theory (RSPT) approach~\cite{Maryasin2014} yields 
$
\g = SJ_{\pm\pm}^3/(192 J_{\pm}^2)
$ from the bare result and 
$
\g \approx 3 J_{\pm\pm}^3/(2000J_{\pm\pm}^2)
$
from a resummation specifically at $S=1/2$. These yield estimates $\g_{\rm RSPT, \rm bare} \approx 0.046\ \mu{\rm eV}$ and $\g_{\rm RSPT, \rm resum} \approx 0.026\ \mu{\rm eV}$. That these estimates are comparable to ours may provide some insight into why we can obtain estimates for $\g$ from such small clusters, given the real space perturbation theory is inherently local and only involves a few neighboring sites. 

\clearpage

\addtolength{\oddsidemargin}{-0.75in}
\addtolength{\evensidemargin}{-0.75in}
\addtolength{\topmargin}{-0.725in}

\newcommand{\addpage}[1] {
\begin{figure*}
  \includegraphics[width=8.5in,page=#1]{supplemental_material.pdf}
\end{figure*}
}

\addpage{1}
\addpage{2}
\addpage{3}
\addpage{4}
\addpage{5}
\addpage{6}
\addpage{7}

\end{document}